\documentclass[11pt,a4paper,twoside,groupcitations]{article}
\usepackage[T1]{fontenc}
\usepackage[ansinew]{inputenc}
\usepackage[english]{babel}
\usepackage{amsfonts}
\usepackage{amsmath}
\usepackage{bm}
\usepackage{array,upgreek,accents}
\usepackage{amsthm}
\usepackage{amssymb}
\usepackage{graphicx}
\usepackage{braket}
\usepackage{eucal}
\usepackage{verbatim}
\usepackage[table]{xcolor}
\usepackage{caption}
\usepackage{cite}
\usepackage{textcomp}
\usepackage{comment}
\usepackage{float}
\usepackage{multicol}
\usepackage{tikz}
\usetikzlibrary{positioning,arrows}
\usetikzlibrary{decorations.pathmorphing}
\usetikzlibrary{decorations.markings}
\usetikzlibrary{calc,decorations.markings}
\usetikzlibrary{arrows,shapes}
\usetikzlibrary{matrix,arrows}

\usepackage[colorlinks,hyperindex,unicode]{hyperref}
\definecolor{green1}{RGB}{0,128,0} 
\hypersetup{hidelinks,
  backref=true,
  pagebackref=true,
  hyperindex=true,
  colorlinks=true,
  breaklinks=true,
urlcolor= blue}
\hypersetup{%
  colorlinks = true,
  linkcolor  = blue,
  citecolor = cyan,
}

\newcommand{\beq}{\begin{eqnarray}}
\newcommand{\eeq}{\end{eqnarray}}
\newcommand{\be}{\begin{eqnarray}}
\newcommand{\ee}{\end{eqnarray}}

\renewcommand{\d}{\mbox{${\rm d}$}} 
\newcommand{\lp}{\ell_{\rm p}}
\newcommand{\mpl}{m_{\rm p}}
\newcommand{\gn}{G_{\rm N}}
\newcommand{\rh}{r_{\rm H}}
\newcommand{\Rh}{R_{\rm H}}

\title{\bf Coherent quantum geometry: de~Sitter spacetime in different foliations}
\author{T.~Bambagiotti$^{ab}$\thanks{E-mail: tommaso.bambagiotti2@unibo.it},
$\, $
R.~Casadio$^{abc}$\thanks{E-mail: casadio@bo.infn.it},
$\, $
A.~Giusti$^{abc}$\thanks{E-mail: A.Giusti@sussex.ac.uk},
$\ $and
P.~Meert$^{d}$\thanks{E-mail: pedro.meert@unesp.br},
\\
\\
$^a${\em Dipartimento di Fisica e Astronomia, Universit\`a di Bologna}
\\
{\em via Irnerio~46, 40126 Bologna, Italy}
\\
\\
$^b${\em I.N.F.N., Sezione di Bologna, I.S.~FLAG}
\\
{\em viale B.~Pichat~6/2, 40127 Bologna, Italy}
\\
\\
$^c${\em Alma Mater Research Center on Applied Mathematics (AM$^2$)}
\\
{\em Via Saragozza 8, 40123 Bologna, Italy}
\\
\\
$^d${\em Instituto de F\'isica Te\'orica, Unesp,}
\\
{\em S\~ao Paulo, 01140-070, Brazil}
}
\begin{document}
\maketitle
\begin{abstract}
In any theory of quantum gravity, an interesting question to address is to what extent known solutions
of the Einstein field equations can be obtained as expectation values of metric operators on suitable
quantum states.
In this work, we consider coherent states (to ensure minimum uncertainty) for different foliations of
the de~Sitter spacetime and study temporal evolution and coordinate invariance.
A general framework will first be introduced for metrics that can be diagonalised globally.
We will then find that the normalisability of coherent states in this framework requires using
reference frames without coordinate singularities.
\end{abstract}
\section{Introduction}
\setcounter{equation}{0}
\label{S:Intro}
Quantum physics involves deviations from classical dynamics that sometimes result in drastically different
behaviours.
A notable example is the stability of atoms, guaranteed by the discrete spectrum of bound states for the electrons:
quantum physics not only avoids the ultra-violet catastrophe of classical electrodynamics, but by doing
so allows the very existence of a macroscopic world.
One could then discuss the same question in the context of gravity and spacetime geometries:
are all solutions of the Einstein theory (or modifications thereof), including those with singularities~\cite{HE},
admissible in a quantum world?
\par
A quantum description for static spherically symmetric geometries based on coherent states 
was introduced for the Schwarzschild spacetime in Ref.~\cite{Casadio:2021eio}.
The main idea was to employ the Fock construction of quantum field theory to build quantum states
as excitations above an absolute vacuum that can reproduce classical solutions of the Einstein equations
as closely as possible. 
For static and spherically symmetric metrics of the form
\be
g_{\mu\nu}\,\d x^\mu\otimes\d x^\nu
=
-A(r)\,\d t\otimes\d t
+
\frac{\d r\otimes\d r}{A(r)}
+
r^2
\left(\d \theta\otimes\d \theta
  +
  \sin^2\theta\,\d \phi\otimes\d \phi
\right)
\ ,
\label{gS}
\ee
with $A=1+2\,V$, a coherent state $\ket{g}$ was obtained such that~\footnote{We shall use
metric signature $(-+++)$ and units with the speed of light $c=1$, the Newton constant $\gn=\lp/\mpl$
and the Planck constant $\hbar=\lp\,\mpl$, where $\lp$ is the Planck length and $\mpl$ the Planck mass.} 
\be
\sqrt{\gn}
\bra{g}\hat \Phi\ket{g}
\simeq
V(r)
\ ,
\label{<V>}
\ee
where $\Phi$ was treated as a canonically quantised massless scalar field in Minkowski spacetime.
This reference Minkowski geometry is naturally associated with the aforementioned absolute vacuum devoid
of physical events in a non-perturbative perspective.~\footnote{The numbers of time and space dimensions
are also purely formal without events.
The physical world emerges from excitations along 3 spatial and 1 time directions.} 
\par
The main outcome of the analysis in Ref.~\cite{Casadio:2021eio} was that possible states $\ket{g}$
in Eq.~\eqref{<V>} are normalisable only if they contain a finite number of excitations with respect
to the vacuum Minkowski geometry.
Imposing a proper normalisation then removes both ultra-violet (UV) and infra-red (IR) singularities
present in the exact classical geometry.
This implies that the Schwarzschild solution cannot be realised exactly and
the allowed states $\ket{g}$ give rise to quantum-corrected geometries
\be
g_{\mu\nu}^{\rm Q}\,\d x^\mu\otimes\d x^\nu
=
-A_{\rm Q}(r)\,\d t\otimes\d t
+
\frac{\d r\otimes\d r}{A{\rm Q}(r)}
+
r^2
\left(\d \theta\otimes\d \theta
  +
  \sin^2\theta\,\d \phi\otimes\d \phi
\right)
\ ,
\label{gQS}
\ee
with $A_{\rm Q}\simeq 1+2\,\sqrt{\gn}\bra{g}\hat \Phi\ket{g}\neq A$,
that appear (effectively or actually~\cite{Casadio:2023ymt}) sourced by matter cores without
inner horizons~\cite{Casadio:2021eio,Feng:2025nai}. 
\par
We remark that no attempt was made in this approach to obtain states associated with macroscopic
configurations, like Eq.~\eqref{gS}, by solving quantum gravity equations~\cite{kiefer}.
In fact, classical geometries are large deformations of the Minkowski vacuum
that we do not expect to be able to reconstruct in general using perturbative expansions or other
known methods. 
On the other hand, the classical geometries we wish to reproduce do satisfy the
equations of General Relativity and we therefore rely on the Ehrenfest theorem
when writing Eq.~\eqref{<V>}.
The same approach was subsequently applied to the Reissner-Nordstr\"om metric~\cite{Casadio:2022ndh}
and the static patches of the de~Sitter (dS) spacetime~\cite{Meert:2025cct} and Schwarzschild-dS
spacetime~\cite{Giusti:2021shf}.
In all cases, similar conclusions were obtained and quantum corrections were derived to the corresponding
line elements from the condition of normalisability of coherent states.
\par
It is important to remark that the condition~\eqref{<V>} preserves the fundamental property of the
(classical) metric to be non-degenerate in the whole range $r>0$ and the quantum-corrected inverse metric
is then given by
\be
g^{\mu\nu}_{\rm Q}\,\partial_\mu\otimes\partial_\nu
=
-\frac{\partial_t\otimes\partial_t}{A_{\rm Q}(r)}
+A_{\rm Q}(r)\,\partial_r\otimes\partial_r
+
\frac{1}{r^2}
\left(
  \partial_\theta\otimes\partial_\theta
  +
  \frac{\partial_\phi\otimes\partial_\phi}{\sin^2\theta}
\right)
\ 
\label{gQA}
\ee
It also follows that, if $A_{\rm Q}(\rh)=0$, the sphere at $r=\rh>0$ is a proper horizon
where the metric is still invertible:
the products $g_{tt}(r)\,g_{rr}(r)=g^{tt}(r)\,g^{rr}(r)=-1$ for all values of $r>0$ and, in particular,
for $r=\rh$~\cite{Jacobson:2007tj}.
\par
Metrics of the form~\eqref{gS} are however very special.
In order to study more general spacetimes and foliations, we are going to consider a scheme
in which all metric components of a given classical geometry in a given reference frame,
\be
\d s^2
=
g_{\mu\nu}\,\d x^\mu\,\d x^\nu
\ ,
\ee
should arise as excitations on the (absolute Minkowki) vacuum of a quantum operator encoded
in a coherent state, namely~\footnote{This relation holds for metric components $g_{\mu\nu}$
that are dimensionless and must be adjusted otherwise.}
\be
\sqrt{\gn}
\bra{g}\hat \Phi_{\mu\nu}\ket{g}
\simeq
g_{\mu\nu}
\ ,
\label{<gmunu>}
\ee
where $\Phi_{\mu\nu}$ are massless (canonically normalised) fields satisfying the vacuum
equation of motion
\be
\Box\Phi_{\mu\nu}
=
0
\ .
\label{eq:KG0}
\ee
In the above, $\Box$ is again the d'Alembert operator in the Minkowski spacetime corresponding to the
absolute vacuum.
Unlike the previous approach, it is important to remark that Eq.~\eqref{<gmunu>} implies that the
classical Minkowski geometry $g_{\mu\nu}=\eta_{\mu\nu}$ should also emerge from a quantum state.
\par
We again expect that well-defined coherent states require UV and/or IR modifications 
in order to yield finite total occupation numbers.
In turn, such modifications will lead to quantum-corrected metrics
\be
A_{\mu\nu}\,\d x^\mu\otimes\d x^\nu
\simeq
\sqrt{\gn}
\bra{g}\hat \Phi_{\mu\nu}\ket{g}
\d x^\mu\otimes\d x^\nu
\ .
\label{g^Q}
\ee
which must be non-degenerate.
However, by also introducing (canonically normalised) scalar fields $\Phi^{\mu\nu}$, satisfying $\Box \Phi^{\mu\nu}=0$
and
\be
B^{\mu\nu}\,\partial_\mu\otimes\partial_\nu
\simeq
\sqrt{\gn}
\bra{g}\hat \Phi^{\mu\nu}\ket{g}
\partial_\mu\otimes\partial_\nu
\ ,
\label{g_Q}
\ee
one does not expect that  
\be
A^{\mu\alpha}\,B_{\alpha\nu}
=
\delta^\mu_\nu
\ .
\ee
The prescription~\eqref{<gmunu>} must therefore be refined in order to obtain a consistent
description of the quantum-corrected spacetime metric.
\par
In the present work, we will define the quantum corrected metric as
\be
g^{Q}_{\mu\nu}
=
\sigma\,\sqrt{\frac{1}{2}\left|(A\,B^{-1}+B^{-1}\,A)_{\mu\nu}\right|}
\ ,
\label{gQ}
\ee
which is symmetric since the matrices $A$ and $B$ are.
Likewise we can define the inverse metric as
\be
g_{Q}^{\mu\nu}
=
\sigma\,\sqrt{\frac{1}{2}\left|(B\,A^{-1}+A^{-1}\,B)^{\mu\nu}\right|}
\ ,
\label{gQ-1}
\ee
which is the inverse of Eq.~\eqref{gQ} only if $A$ and $B$ commute, that is if they can be
simultaneously diagonalised.
In the above, $\sigma=\pm 1$ must be chosen so as to preserve the Lorentzian metric signature,
a crucial point that will be discussed explicitly later on.
Moreover, since Eq.~\eqref{<gmunu>}, like Eq.~\eqref{<V>}, is defined in a specific reference frame,
other questions that we will start to address are whether the states $\ket{g}$ require using specific
reference frames and to what extent general coordinate invariance is preserved.
\par
In order to put the above approach to the test, we shall apply it to the spatially flat de~Sitter (dS)
spacetime~\cite{HE}.
We recall that this manifold can be foliated in different ways and we shall in particular consider
the synchronous coordinate system in which the metric is given by
\be
\d s^{2}
&\!\!=\!\!&
-\d t^{2}
+e^{2\,H\,t}\,
\sum_{i=1}^3(\d x^i)^2
\label{ds_xyz}
\\
&\!\!=\!\!&
-\d t^{2}
+e^{2\,H\,t}\left(\d r^{2}
  +
r^{2}\d\Omega^{2}\right)
\ ,
\label{eq:ds-cscs}
\ee
where $ H=\sqrt{\Lambda/3}$ and $\Lambda>0$ is the cosmological constant,
$x^i$ (with $i=1,2,3$) are orthonormal coordinates on flat hypersurfaces of constant $t$ and
$(r,\theta,\phi)$ the corresponding spherical coordinates. 
The above reference frames cover the so-called expanding Poincar\'e patch of the whole dS manifold
which is of particular interest in cosmology.
The Poincar\'e patch includes the smaller static patch in which the line element reads
\be
\d s^{2}
=
-
\left(1-H^2\,R^{2}\right)\d T^{2}
+
\frac{\d R^{2}}{\left(1-H^2\,R^{2}\right)}
+
R^{2}\,\d\Omega^{2}
\ .
\label{eq:ds-static}
\ee
One can relate the coordinates in the two foliations using the transformation~\footnote{A
recent review of these coordinate systems can be found~in Ref.~\cite{Nouri-Zonoz:2014tla}.}
\be
\left\{
  \begin{aligned}
    t & =T+\frac{\ln\!\left(1-H^2\,R^{2}\right)}{2\,H}
    \\
    r & =\frac{R\,e^{-H\,T}}{\sqrt{1-H^2\,R^{2}}}
    \ .
  \end{aligned}
\right.
\label{eq:coordtransformation}
\ee
\par
For metrics that are already in diagonal form globally, like those in Eq.~\eqref{ds_xyz} or \eqref{eq:ds-static},
the prescription~\eqref{gQ} yields (no sum over the index $\mu$)
\be
\left(g_{\mu\mu}^{\rm Q}\right)^2
=
\frac{\bra{g}\hat \Phi_{\mu\mu}\ket{g}}{\bra{g}\hat \Phi^{\mu\mu}\ket{g}}
\label{gQdd}
\ee
and the inverse~\eqref{gQ-1} reads
\be
\left(g^{\mu\mu}_{\rm Q}\right)^2
=
\frac{\bra{g}\hat \Phi^{\mu\mu}\ket{g}}{\bra{g}\hat \Phi_{\mu\mu}\ket{g}}
\ ,
\label{gQuu}
\ee
which ensure that 
\be
\left(g^{\mu\mu}_{\rm Q}\right)^2
\left(g_{\mu\mu}^{\rm Q}\right)^2
=
1
\ .
\ee
We remark once more that particular care will have to be taken in choosing the sign of the components
(see Section~\ref{S:static}).
\par
In Section~\ref{S:CSCS}, we shall build a coherent state that reproduces the line elements~\eqref{ds_xyz}
and~\eqref{eq:ds-cscs} as closely as possible.
We will see that the condition of normalisability will still result in the classical metric in the synchronous foliation,
albeit with a restriction on the (maximum value of the) time variable.
In Section~\ref{S:static}, we will try to build a coherent state for the metric directly in the static patch.
In that case, however, we will see that normalisable coherent states necessarily replace the horizon
with a real singularity. 
\section{Poincar\'e patch}
\setcounter{equation}{0}
\label{S:CSCS}
We start our analysis from the dS metric in the synchronous coordinates that cover the expanding 
Poincar\'e patch.
We will build coherent states that reconstruct the metrics in the forms~\eqref{ds_xyz}
and~\eqref{eq:ds-cscs}, and show that these constructions are fully equivalent.
\subsection{Coherent state for Cartesian coordinates}
\label{SS:cartesian}
Adopting cartesian coordinates $x^\mu=(t,x^i)\equiv (t,\vec x)$ for the Minkowski metric,
the Klein-Gordon Eq.~\eqref{eq:KG0} becomes~\footnote{We omit Greek indices when convenient.
We also assume $\Phi^2$ has canonical dimensions of mass over length.}
\be
\left[
-\frac{\partial^2}{\partial t^2}
+
\sum_{i=1}^3
\frac{\partial^2}{\partial (x^i)^2}
\right]
\Phi(t,\vec x)
=
0
\ .
\label{KGmink}
\ee
The normal modes are simply given by plane waves,
\be
u_{\vec{k}}
=
\frac{e^{-i\,k\,t}}{\sqrt{2\,k}}
\,e^{i\,k_{j}\,x^{j}}
\ ,
\ee
where $k=\sqrt{k_i\,k^i}$, which satisfy the orthogonality relation in the Klein-Gordon scalar
product
\be
i
\int
\frac{\d^{3}x}{\left(2\,\pi\right)^{3}}\,
u^{*}_{\vec{k}}(t,\vec{x})\,\overset{\leftrightarrow}{\partial}_{t}\,u_{\vec{p}}(t,\vec{x})
&\!\!=\!\!&
\prod^{3}_{j=1}
\int\limits^{+\infty}_{-\infty}
\frac{\d x^{j}}{2\,\pi}\,
e^{-i\,(k_{j}-p_{j})\,x^{j}}
\nonumber
\\
&\!\!=\!\!&
\prod^{3}_{j=1}
\delta(k_{j}-p_{j})
\equiv
\delta_{3}(\vec{k}-\vec{p})
\ .
\label{eq:ortog1}
\ee
The field operators are written as
\be
\hat{\Phi}(t,\vec x)
=
\int
\d^3 k\,\sqrt{\hbar}
\left[u_{\vec k}(t,\vec x)\,\hat{a}_{\vec k}
+
u_{\vec k}^{*}(t,\vec x)\,\hat{a}_{\vec k}^{\dagger}
\right]
\ee
and $\hat{\Pi}(t,\vec{x})=\partial_{t}\hat{\Phi}(t,\vec{x})$.
Equal time commutation relations $\left[\hat{\Phi}(t,\vec{x}),\hat{\Pi}(t,\vec{y})\right]=i\,\hbar\,\delta(\vec{x}-\vec{y})$
then lead to the standard commutators for annihilation and creation operators
\be
\left[
\hat{a}_{\vec k}^{\phantom\dagger},\hat{a}_{\vec p}^{\dagger}
\right]
=
\delta_{3}(\vec k-\vec p)
\ .
\ee
\par
Coherent states are eigenstates of the annihilation operators,
\be
\hat a_{\vec k}\ket{g}
=
e^{i\,\gamma_k(t)}\,
g_{\vec k}
\ket{g}
\ee
and can be written as
\be
\ket g
=
e^{-\frac{\mathcal N}{2}}
\exp\left\{
\displaystyle
\int
\d^3 k\,
g_{\vec k}\,\hat{a}_{\vec k}^{\dagger}
\right\}
\ket 0
\ ,
\label{eq:gstateC}
\ee
where the normalisation constant (or total occupation number) reads
\be
\mathcal{N}
=
\int
\d^3 k
\left|g_{\vec k}\right|^{2}
\ .
\label{eq:NC}
\ee
This condition generically imposes deviations from the classical behaviour, either in the UV (that is,
modified $ g_{\vec k}$ for $k\to \infty $) or in the IR (modified $g_{\vec k}$ for $k\to 0 $), or both.
Such deviations resulted in quantum-corrected line elements in
Refs.~\cite{Casadio:2021eio,Feng:2025nai,Casadio:2022ndh,Meert:2025cct,Giusti:2021shf}.
\par
We then find $\gamma_k=k\,t$ and
\be
g_{\vec k}
=
\frac{\sqrt{2\,k}}{\lp}\,
\int
\frac{\d^3 x}{(2\,\pi)^3}\,
e^{-i\,k_i\,x^i}\,f(t)
=
\frac{\sqrt{2\,k}}{\lp}\,
f(t)\,\delta_3(\vec k)
\ ,
\ee
where $f=-1$ for $g_{tt}$ and $f=e^{2\,H\,t}$ for $g_{ii}$. 
In order for the normalisation~\eqref{eq:NC} to be finite, we regularise the Dirac delta distributions,
\be
\delta(k_i)
\to
\frac{e^{-{k_i^2}/{\sigma_i^2}}}{\sqrt{\pi}\,\sigma_i}
\ ,
\label{deltaReg}
\ee
where we could employ different IR scales $\sigma_i$ along the coordinates $x^i$.
This regularisation yields
\be
\sqrt{\gn}
\bra{g}\hat \Phi_{tt}\ket{g}
=
-e^{-\frac{\sum_i\sigma_i^2\,(x^i)^2}{4}}
\ee
and
\be
\sqrt{\gn}
\bra{g}\hat \Phi_{ii}\ket{g}
=
e^{-\frac{\sum_j\sigma_j^2\,(x^j)^2}{4}}
\,
e^{2\,H\,t}
\ .
\ee
The total occupation numbers~\eqref{eq:NC} become
\be
\mathcal N_{tt}
=
\frac{2}{\lp^2}
\int
\frac{\d^3 k}{\pi^3}\,
\sqrt{\sum_j k_j^2}\,
\frac{e^{-2\,\sum_i\,k_i^2/\sigma_i^2}}
{\prod_j\sigma_j^2}
\ ,
\label{NttPC}
\ee
and 
\be
\mathcal N_{ii}=\mathcal N_{tt}\,e^{4\,H\,t}
\ ,
\label{NiiPC}
\ee
which are now finite for $t<\infty$.
Moreover, $\mathcal N_{tt}$ is constant (like it would be in Minkowski)
but $\mathcal N_{ii}$ vanishes for $t\to -\infty$.
For isotropic regularisation, with $\sigma_i=\sigma$, one can easily find
\be
\mathcal N_{tt}
=
\frac{8}{\lp^2}
\int\limits_0^\infty
\frac{k^3\,\d k}{\pi^2\,\sigma^6}\,
e^{-2\,k^2/\sigma^2}
=
\frac{1}{\pi^2\,\lp^2\,\sigma^2}
\ .
\label{N_PC}
\ee
For instance, on assuming $\sigma\sim H$, we obtain a Bekenstein-like
counting~\cite{Bekenstein:1973ur,Dvali:2013eja}.
\par
According to the prescription~\eqref{g^Q}, the quantum-corrected version of the
metric~\eqref{ds_xyz} then reads
\be
A_{\mu\nu}\,\d x^\mu\,\d x^\nu
=
e^{-\frac{\sum_i\,\sigma_i^2\,(x^i)^2}{4}}
\left[
  -\d t^{2}
  +e^{2\,H\,t}\,
  \sum_{i=1}^3
  (\d x^i)^2
\right]
\ ,
\label{eq:ds-xyzQ}
\ee
which shows that both homogeneity and isotropy are lost in general.
However, isotropy can be restored  for $\sigma_i=\sigma$ and $r^2=\sum_{i=1}^3 (x^i)^2$. 
In the following, we shall assume a regularisation that preserves isotropy.
\par
The matrix~\eqref{eq:ds-xyzQ} is clearly regular in the whole Poincar\'e patch
and its inverse is given by
\be
(A^{-1})^{\mu\nu}\,\partial_\mu\otimes\partial_\nu
=
e^{+\frac{\sigma^2\,r^2}{4}}
\left[
  -\partial_t\otimes\partial_t
  +e^{-2\,H\,t}\,
  \sum_{i=1}^3
  \partial_i\otimes\partial_i
\right]
\ .
\ee
However, if we compute the inverse metric directly from a coherent state, we obtain
\be
\sqrt{\gn}
\bra{g}\hat \Phi^{tt}\ket{g}
=
-e^{-\frac{\sigma^2\,r^2}{4}}
\ee
and
\be
\sqrt{\gn}
\bra{g}\hat \Phi^{jj}\ket{g}
=
e^{-\frac{\sigma^2\,r^2}{4}}\,
e^{-2\,H\,t}
\ ,
\ee
so that
\be
B^{\mu\nu}\,\partial_\mu\otimes\partial_\nu
=
e^{-\frac{\sigma^2\,r^2}{4}}
\left[
  -\partial_t\otimes\partial_t
  +e^{-2\,H\,t}\,
  \sum_{i=1}^3
  \partial_i\otimes\partial_i
\right]
\ .
\ee
This simple case shows explicitly the problem with the prescription~\eqref{<gmunu>}
that we anticipated in the introductory Section~\ref{S:Intro}.
\par
By applying Eqs.~\eqref{gQdd} and~\eqref{gQuu}, we now find
\be
g_{tt}^{\rm Q}
=
-1
=
g_{tt}
\ee
and
\be
g_{ii}^{\rm Q}
=
e^{2\,H\,t}
=
g_{ii}
\ ,
\ee
so that $\d s_{\rm Q}^{2}=\d s^2$.
Homogeneity is restored by the new prescription and the dS metric does not acquire any quantum
corrections in the Poincar\'e patch in Cartesian coordinates.
In fact, one could argue that the same result should be obtained from the old
prescription~\eqref{<V>} since $V(r)=0$ in the Poincar\'e patch.
However, we remark that the above construction holds for $\mathcal N_{\mu\nu}$ finite.
From Eq.~\eqref{NiiPC}, we must therefore have $t<\infty$, in agreement with
analyses suggesting that the dS spacetime is unstable in quantum
gravity~\cite{Woodard:2004ut,Ho:2015bua,Rajaraman:2016nvv,Dvali:2017eba}.
\subsection{Coherent state for spherical coordinates}
\label{SS:sphere}
We shall now repeat the analysis of the Poincar\'e patch in spherical coordinates to
test how robust the previous conclusions are with respect to general covariance,
albeit in a frame that still allows for explicit calculations.
\par
In spherical coordinates, Eq.~\eqref{KGmink} reads 
\be
\Box\Phi
=
\left[
  -
  \frac{\partial^2}{\partial t^2}
  +
  \frac{1}{r^{2}}\,\frac{\partial}{\partial r}
  \left(r^{2}\,\frac{\partial}{\partial r}\right)
  +
  \frac{1}{r^{2}\sin\theta}\,\frac{\partial}{\partial \theta}
  \left(\sin\theta\,\frac{\partial}{\partial \theta}\right)
  +
  \frac{1}{r^{2}\sin^{2}\theta}
  \frac{\partial^2}{\partial \phi^2}
\right]\Phi(t,r) 
=
0
\ .
\label{eq:KGfull}
\ee
The metric operators can be written in this reference frame as
\be
\hat{\Phi}(t,r)
=
\int_0^\infty
\frac{k^{2}\,\d k}{2\,\pi^{2}}\,
\sqrt{\frac{\hbar}{2\,k}}
\left[u_{k}(t,r)\,\hat{a}_{k}
+u_{k}^{*}(t,r)\,\hat{a}_{k}^{\dagger}
\right]
\ee
and their conjugate momenta $\hat{\Pi}=\partial_{t}\hat{\Phi}$, where 
the normal modes
\be
u_{k}(t,r)
=
e^{-\,i\,k\,t}\,j_{0}(k\,r)
\ ,
\label{u_k}
\ee
are given by the spherical Bessel functions of order zero,
\be
j_{0}(k\,r)
=
\frac{\sin(k\,r)}{k\,r}
\ ,
\ee
satisfying the orthogonality relations
\be
4\,\pi
\int_{0}^{\infty}
r^{2}\,\d r\,
j_0(k\,r)\,j_0(p\,r)
=
\frac{2\,\pi^{2}}{k^{2}}\,
\delta(k-p)
\label{eq:bessel_ortog}
\ee
and
\be
\int_{0}^{\infty}
\frac{k^{2}\,\d k}{2\,\pi^{2}}\,
j_{0}\left(k\,r\right)j_{0}\left(k\,s\right)
=
\frac{1}{4\,\pi\, r^{2}}\,\delta(r-s)
\ .
\ee
Eq.~\eqref{eq:bessel_ortog} then implies the usual equal-time commutation rules
\be
\left[
\hat{\Phi}(t,r),\hat{\Pi}(t,s)
\right]
=
\frac{i\,\hbar}{4\,\pi\,r^{2}}\,\delta(r-s)
\quad
\iff
\quad
\left[
\hat{a}_{k},\hat{a}_{p}^{\dagger}
\right]
=
\frac{2\,\pi^{2}}{k^{2}}\,\delta(k-p)
\ ,
\label{eq:OP_com}
\ee
so that $\hat a_{k}$ and $\hat a_{k}^{\dagger}$ are annihilation and creation operators, respectively.
\par
The Fock space construction proceeds like in Section~\ref{SS:cartesian} and
the coherent states~\eqref{eq:gstateC} are now determined from the expectation values
\be
\sqrt{\gn}
\bra{g}\hat{\Phi}_{\mu\nu}\ket{g}
=
\sqrt{\frac{\lp}{\mpl}}
\int_0^\infty
\frac{k^{2}\,\d k}{2\,\pi^{2}}\,
\sqrt{\frac{2\,\lp\,\mpl}{k}}\,g_k\,j_0(k\,r)\,\cos(k\,t-\gamma_k)
=
g_{\mu\nu}(t,r)
\ ,
\label{eq:Phi_expct}
\ee
which should hold along with the normalisability of $\ket{g}$. 
As in the case of Cartesian coordinates, this condition will also impose deviations from classical behaviour
in the form of a regularisation.
\par
The coherent state $\ket g$ must reproduce the metric~\eqref{eq:ds-cscs}, that is
\be
g_{tt}
=
-1
\ ,
\quad
g_{rr}
=
e^{2\,H\,t}
\ ,
\quad
g_{\theta\theta}
=
e^{2\,H\,t}\,r^2
\ .
\label{eq:gCSCS}
\ee
Note in particular that we must then have $A_{tt}=\sqrt{\gn}\bra{g} \hat \Phi_{\theta\theta}\ket{g}$,
$A_{rr}=\sqrt{\gn}\bra{g} \hat \Phi_{\theta\theta}\ket{g}$ but
$A_{\theta\theta}=\lp^2\,\sqrt{\gn}\bra{g} \hat \Phi_{\theta\theta}\ket{g}$
for dimensional consistency.
The relevant coefficients are computed in Appendix~\ref{A:useful1} and take the general form
\be
g_k\,e^{i\,\gamma_k}
=
\sqrt{\frac{k}{2}}\,
\frac{\tilde f_k^{(\alpha)}[f]}
{\lp^{1+\alpha}}
\ ,
\ee
with $\tilde f_k^{(\alpha)}[f]$ given in Eq.~\eqref{tr2}.
We then find $\gamma_k=k\,t$ and
\be
g_k
=
-\frac{4\,\pi^2\,f(t)}{\sqrt{2\,k}\,\lp^{1+\alpha}\,i^\alpha}\,
\delta^{(1+\alpha)}(k)
\ ,
\label{gkp}
\ee
where $ \delta^{(\alpha+1)} $ indicates the $\left(\alpha+1\right)^{\text{th}} $-derivative of the
(1-dimensional) delta distribution.
Each component of $ A_{\mu\nu} $ is associated with a different $f( t)$ and
value of $ \alpha $, for instance $f(t)=-1$ with $\alpha=0$ for $A_{tt}$, $f(t)=e^{2\,H\,t}$
with $\alpha=0$ for $A_{rr}$, and $f(t)=e^{2\,H\,t}$ with $\alpha=2$ for $A_{\theta\theta}$.
\par
The total occupation number~\eqref{eq:NC} for each component now reads
\be
\mathcal{N}
=
\frac{4\,\pi^2\,f^2(t)}{\lp^{2\,(1+\alpha)}}
\int_0^\infty
k\,\d k
\left[\delta^{(1+\alpha)}(k)\right]^2
\ ,
\ee
which is not well defined (in the IR, $k\to 0$).
As $ g_{k}\propto\delta^{\left(\alpha+1\right)}\left(k\right) $, we apply the same regularisation scheme for
the delta distribution as in Eq.~\eqref{deltaReg}, but due to spherical symmetry we have
\be
\delta(k)
\to
\frac{e^{-\frac{k^2}{\sigma^2}}}{\sqrt{\pi}\,\sigma}
\ ,
\ee
where $\sigma\sim 1/R_\infty>0$ acts as an IR cut-off scale associated with the spatial volume
proportional to $R_\infty^3$.
This yields
\par
\be
\delta^{(1+\alpha)}(k)
\to
\left\{
  \begin{array}{ll}
    -\strut\displaystyle\frac{2\,k\,e^{-\frac{k^2}{\sigma^2}}}{\sqrt{\pi}\,\sigma^3}
&
\quad 
{\rm for}\ \alpha=0
\\
\\
-\strut\displaystyle4\,k\,\frac{(2\,k^2-3\,\sigma^2)\,e^{-\frac{k^2}{\sigma^2}}}{\sqrt{\pi}\,\sigma^7}
&
\quad 
{\rm for}\ \alpha=2
\ .
  \end{array}
\right.
\ee
For $\alpha=0$, we have
\be
\mathcal{N}
=
\frac{16\,\pi\,f^2(t)}{\lp^2\,\sigma^6}
\int_0^\infty
k^3\,\d k \,
e^{-\frac{2\,k^2}{\sigma^2}}
=
\frac{2\,\pi\,f^2(t)}{\lp^2\,\sigma^2}
\ ,
\label{Ncscs1}
\ee
Likewise, for $\alpha=2$, we obtain
\par
\be
\mathcal{N}
=
\frac{64\,\pi\,f^2(t)}{\lp^6\,\sigma^{14}}
\int_0^\infty
k^3 \,\d k \,
(2\,k^2-3\,\sigma^2)^2\,
e^{-\frac{2\,k^2}{\sigma^2}}
=
\frac{24\,\pi\,f^2(t)}{\lp^6\,\sigma^6}
\ .
\label{Ncscs2}
\ee
In general, all $\mathcal N$ display the behaviour
\be
\mathcal N
\propto
\frac{f^2(t)}{\left(\lp^{2}\,\sigma^{2}\right)^{1+\alpha}}
\ ,
\label{eq:sphericalOccN}
\ee
meaning that for $ \bra g \hat{\Phi}_{tt} \ket g $ it is finite for all $t$, whereas it diverges for
$\bra g \hat{\Phi}_{rr} \ket g$ and $ \bra g \hat{\Phi}_{\theta\theta} \ket g$ in the limit $t\to\infty$,
in agreement with the results in Cartesian coordinates from Section~\ref{SS:cartesian}.
\par
Using the expressions~\eqref{eq:regQC1} and~\eqref{eq:regQC2}, we now find
\be
\sqrt{\gn}
\bra{g}\hat \Phi_{\mu\nu}\ket{g}
=
-e^{{-\frac{\sigma^2\,r^2}{4}}}\,
g_{\mu\nu}
\ ,
\ee
where $g_{\mu\nu}$ denote again the classical components in Eq.~\eqref{eq:ds-cscs}.
One therefore obtains
\be
A_{\mu\nu}\,\d x^\mu\,\d x^\nu
=
e^{{-\frac{\sigma^2\,r^2}{4}}}
\left[
-\d t^{2}
+e^{2\,H\,t}
\left(\d r^{2}
+
r^{2}\,\d\Omega^{2}\right)
\right]
\ ,
\label{dSq}
\ee
so that the quantum-corrected metric determined by the regularised coherent state 
could have been obtained from the result~\eqref{eq:ds-xyzQ} by the standard change
from Cartesian coordinates to spherical coordinates.
All of the conclusions drawn in Section~\ref{SS:cartesian} are therefore confirmed.
The only difference is that the occupation numbers $\mathcal N_{rr}$ and $\mathcal N_{\theta\theta}$
differ.
\section{Static patch}
\setcounter{equation}{0}
\label{S:static}
In the previous Section, we showed that the dS metric does not acquire any quantum corrections
in the Poincar\'e patch if we employ the prescription in Eqs.~\eqref{gQdd} and~\eqref{gQuu}.
The change of coordinates~\eqref{eq:coordtransformation} would therefore lead to the
classical metric~\eqref{eq:ds-static} in the static patch.
In Ref.~\cite{Meert:2025cct} a quantum-corrected metric was obtained by applying the prescription
in Eq.~\eqref{<V>} in the static patch, which yields
\be
\d s_{\rm Q}^2
=
-\left(1-H^2\,R^2\,e^{-\frac{\sigma^2\,R^2}{4}}\right)\d T^2
+
\left(1-H^2\,R^2\,e^{-\frac{\sigma^2\,R^2}{4}}\right)^{-1}\d R^2
+
R^2\,\d\Omega^2
\ .
\label{dsQold}
\ee
We now want to see if the new prescription also entails a quantum-corrected metric.
We will see that the presence of a coordinate singularity at $R=H^{-1}$ makes a crucial difference.
\subsection{Coherent state}
%
%
%
%
%
\label{SS:QmetricStatic}
We next wish to find a coherent state and apply the prescription~\eqref{gQdd} and~\eqref{gQuu}
to define the quantum-corrected metric.
However, we notice that Eqs.~\eqref{gQdd} and~\eqref{gQuu} do not in general ensure that the product
$g^{\rm Q}_{TT}(\Rh)\,g^{\rm Q}_{RR}(\Rh)\sim 1$ for $R=\Rh$ where $g^{\rm Q}_{TT}(\Rh)=0$. 
For instance, let us assume that 
\be
\bra{g}\hat{\Phi}_{TT}\ket{g}
\sim 
(R-\Rh)^a
\ ,
\label{phiTTalpha}
\ee
with $a>0$, for $R\sim\Rh$ and define $\bra{g}\hat{\Phi}_{RR}\ket{g}\equiv \Gamma(R)$.
We then have $\bra{g}\hat{\Phi}^{TT}\ket{g}= \Gamma(R)$ and $\bra{g}\hat{\Phi}^{RR}\ket{g}\sim (R-\Rh)^a$,
from which
\be
\left(g^{\rm Q}_{TT}\right)^2
\sim
\frac{|R-\Rh|^a}{ \Gamma(R)}
\ee 
and
\be
\left(g^{\rm Q}_{RR}\right)^2
\sim
\frac{ \Gamma(R)}{|R-\Rh|^a}
\ .
\ee 
It follows that $g^{\rm Q}_{TT}(\Rh)=0$ only if $\Gamma(R)\sim (R-\Rh)^b$ with $b<a$ for $R\sim\Rh$.
\par
In principle, one might also have 
\be
\bra{g}\hat{\Phi}_{RR}\ket{g}\sim (R-R_0)^b
\ ,
\ee
with $b>0$.
Let us then denote $\bra{g}\hat{\Phi}_{TT}\ket{g}\equiv \Pi(R)$, so that
\be
\left(g^{\rm Q}_{TT}\right)^2
\sim
\frac{\Pi(R)}{|R-R_0|^b}
\ .
\ee 
and
\be
\left(g^{\rm Q}_{RR}\right)^2
\sim
\frac{|R-R_0|^b}{\Pi(R)}
\ .
\ee 
It is then clear that $R_0$ is a real singularity, unless $\Pi(R)\sim(R-R_0)^a$ with $a>b$,
in which case we recover $R_0=\Rh$.
\par
From Eq.~\eqref{eq:gttQCstatic}, we immediately obtain
\be
\sqrt{\gn}
\bra{g}\hat{\Phi}_{TT}\ket{g}
=
-\left(1-H^2\,R^2\right)
e^{-\frac{R^2\,\Sigma^2}{4}}
\ ,
\ee
which is of the form in Eq.~\eqref{phiTTalpha} with $\Rh=1/H$ and $a=1$.
It follows that we must have
\be
\bra{g}\hat{\Phi}_{RR}\ket{g}\sim (R-\Rh)^b
\ ,
\ee
with $b<1$ and no other zeros $R_0$ of $\bra{g}\hat{\Phi}_{RR}\ket{g}$.
\par
\begin{figure}[t]
  \centering
   \includegraphics[width=0.5\linewidth]{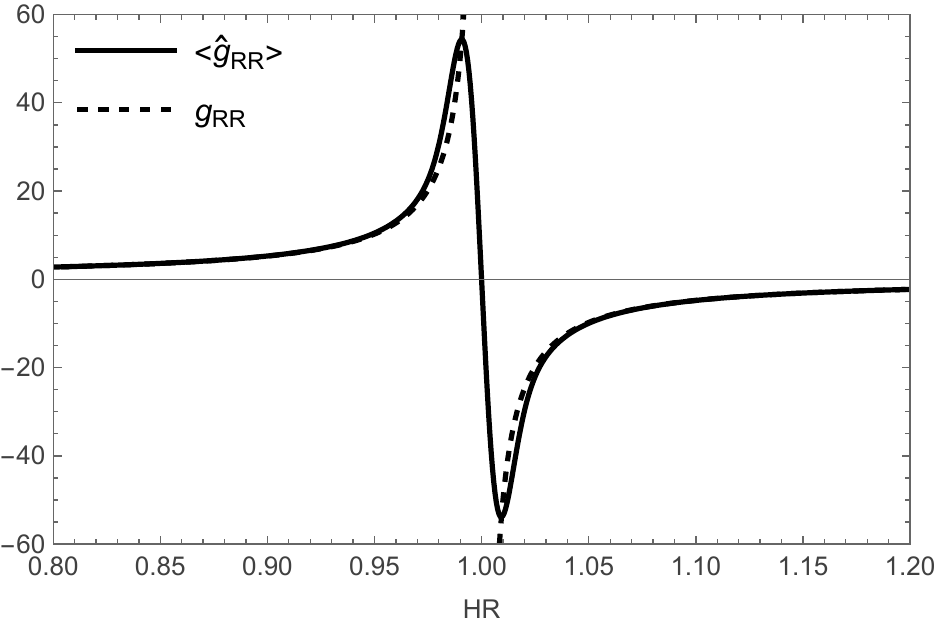}
  \caption{Function $\bra{g}\hat{\Phi}_{RR}\ket{g}$ with the Gaussian UV regularisation
  compared to classical $g_{RR}$.}
  \label{fig:gRRb}
\end{figure}

From Eq.~\eqref{eq:grrQCgauss}, we obtain
\be
\sqrt{\gn}
\bra{g}\hat{\Phi}_{RR}\ket{g}
=
-\frac{1}{H^2\,\ell\,R}
\left[D_+\!\left(\frac{H\,R-1}{H\,\ell}\right)
  +
  D_+\!\left(\frac{H\,R+1}{H\,\ell}\right)
\right]
\ ,
\label{eq:grrQCgaussM}
\ee
which vanishes for $R_0<\Rh=1/H$, since
\be
\bra{g}\hat{\Phi}_{RR}(R=1/H)\ket{g}
=
\frac{1}{H\ell}\left[D_{+}(0)+D_{+}\!\left(\frac{2}{H\,\ell}\right)\right]
=
\frac{1}{H\ell}\,D_{+}\!\left(\frac{2}{H\,\ell}\right)
\ ,
\ee
which is finite and negative for $H\,\ell\ll 1$,
\be
\lim_{H\,\ell\to 0}
\frac{1}{H\ell}\,D_{+}\!\left(\frac{2}{H\,\ell}\right)
=
-\frac{1}{4}
\ .
\ee
Fig.~\ref{fig:gRRb} clearly displays the problem: the classical component $g_{RR}$
shows a discontinuity at $R=\Rh=1/H$ where it diverges positively (negatively) for $R\to \Rh^-$
(for $R\to \Rh^+$).
Any regularised approximation $\bra{g}\hat{\Phi}_{RR}\ket{g}$ will go continuously from
a maximum for $R<\Rh$ to a minimum for $R>\Rh$, which implies the existence of a zero
at $R=R_0$ (in general different from $\Rh$).~\footnote{By employing a sharp UV cut-off,
it is straightforward to see that the discontinuous behaviour around $R=\Rh$ can only be
recovered by including modes of arbitrarily large momentum $k$ (arbitrarily short wavelength
around $R=\Rh$), which result in a diverging total occupation number.}
The conclusion is therefore that the classical causal structure (Hubble horizon) cannot be 
reproduced without introducing a singularity at $R=R_0$.
\par
General covariance for the dS spacetime seems to be drastically lost at the quantum level
(Fock space) in the present framework defined by Eqs.~\eqref{gQ} and~\eqref{gQ-1} in reference
frames where coordinate singularities are present.
Let us further note that a milder breaking occurs also with the old prescription~\eqref{<V>}:
since $V=0$ in the Poincar\'e patch, one should find no quantum corrections to the metric~\eqref{eq:ds-cscs},
whereas the metric~\eqref{eq:ds-static} acquires quantum corrections in the static patch shown
in Eq.~\eqref{dsQold}, albeit without the emergence of a real singularity~\cite{Meert:2025cct}.
\section{Conclusions}
\setcounter{equation}{0}
\label{S:conc}
In this work we have investigated some fundamental aspects associated with the use of coherent states
for reconstructing the dS spacetime.
The choice was motivated by the fact that different foliations display features that had not been considered previously.
First, we notice that the Poincar\'e patch being time-dependent requires the inclusion of time dependence
in the coherent states.
This generalises upon previous works in which only static configurations had been
considered~\cite{Casadio:2021eio,Feng:2025nai,Casadio:2022ndh,Meert:2025cct,Giusti:2021shf}.
In doing so, we noticed that a more general prescription was required to define the quantum-corrected
metric components from expectation values on the coherent states, leading to Eqs.~\eqref{gQ} and~\eqref{gQ-1}.
\par
The proposed new prescription applied to the dS spacetime showed that the emergent metric is sensible
to the choice of foliation (through the correspondingly adapted coordinates) for the classical manifold.
We showed that the occupation numbers associated with the Poincar\'e patch~\eqref{ds_xyz}-\eqref{eq:ds-cscs}
remain finite and the coherent states that reproduce the dS geometry are well-defined only for cosmological time
intervals that are bounded above. 
This implies that eternal dS cannot be realised in quantum gravity, in agreement with different
approaches~\cite{Woodard:2004ut,Ho:2015bua,Rajaraman:2016nvv,Dvali:2017eba}).
We explicitly constructed the coherent states using Cartesian and spherical coordinates and
showed that the two reference frames lead to the same metric with components related by the classical
change of coordinates.
\par
In order to test general covariance on more general grounds, we applied the generalised
prescriptions~\eqref{gQ} and~\eqref{gQ-1} to the static patch~\eqref{eq:ds-static} within the
cosmological dS horizon.
In this case, the total occupation numbers must be regulated both in the IR (like previously
found in Ref.~\cite{Meert:2025cct}), and in the UV.
However, removing the UV divergences inevitably induces a real singularity in the quantum-corrected
metric near the coordinate singularity.
This means that the prescriptions~\eqref{gQ} and~\eqref{gQ-1} can only be applied in 
reference frames that are smooth across horizons.
\par
Another basic restrictions of Eqs.~\eqref{gQ} and~\eqref{gQ-1} is that it only applies to metrics
that can be put in diagonal form globally, hence leading to Eqs.~\eqref{gQdd} and~\eqref{gQuu}. 
Of course, one might wonder if there are different prescriptions that do not involve such restrictions
and still preserve the fundamental properties that we require of the reconstructed spacetime metrics.
For instance, one could consider a local, or patchwork-like, reconstruction of the geometry.
Such questions are left for future investigations.
\subsection*{Acknowledgements}
T.B., R.C.~and A.G.~are partially supported by the INFN grant FLAG.
A.G.~is supported by the Italian Ministry of Universities and Research (MUR)
through the grant ``BACHQ: Black Holes and The Quantum'' (grant no.~J33C24003220006). 
R.C.~and A.G.~carried out this work in the framework of the activities of the National Group
of Mathematical Physics (GNFM, INdAM).
P.M.~is supported by the S\~ao Paulo Research Foundation (FAPESP), Grant no.~2022/12401-9.
\appendix
\section{Poincar\'e patch in spherical coordinates}
\setcounter{equation}{0}
\label{A:useful1}
The coefficients that define the coherent state that reproduce the classical metric component
$g_{\mu\nu}$ can be expressed as
\be
g_{k}\,e^{i\,\gamma_k}
=
\sqrt{\frac{k}{2}}\,\frac{\tilde f_k}{\lp}
\ ,
\label{g_k}
\ee
where 
\be
\tilde f_k(t)
=
4\,\pi
\int_0^\infty
r^2\,\d r\,j_0(k\,r)\,e^{i\,k\,t}\,g_{\mu\nu}(t,r)
\ .
\ee
\subsection{Poincar\'e patch}
In the synchronous coordinates, we need to reproduce the functions in Eq.~\eqref{eq:gCSCS},
which are of the form $f(t)\,r^{\alpha}$, where $f=-1$ with $\alpha=0$ or $f=e^{2\,H\,t}$ with
$\alpha=0$ or $\alpha=2$.
We immediately obtain
\be
\tilde f_k^{(\alpha)}[f]
&\!\!=\!\!&
4\,\pi
\int_0^\infty
r^2\,\d r\,j_0(k\,r)\,e^{i\,k\,t}\,f(t)\,r^\alpha
\nonumber
\\
&\!\!=\!\!&
4\,\pi\,f(t)\,
e^{i\,k\,t}
\int_0^\infty
r^{2+\alpha}\,\d r\,j_0(k\,r)
\nonumber
\\
&\!\!=\!\!&
\frac{2\,\pi}{i\,k}\,f(t)\,e^{i\,k\,t}
\int_{-\infty}^{+\infty}
r^{1+\alpha}\,\d r\,
e^{i\,k\,r}
\nonumber
\\
&\!\!=\!\!&
-\frac{4\,\pi^2}{i^\alpha\,k}\,f(t)\,e^{i\,k\,t}\,
\delta^{(1+\alpha)}(k)
\ ,
\label{tr2}
\ee
where $\delta^{(n)}$ denotes the $n^{\rm th}$ derivative of the Dirac delta distribution.
We then find $\gamma_k=k\,t$ and $g_k$ in Eq.~\eqref{gkp}.
We can next check that we indeed obtain the expected function from Eq.~\eqref{eq:Phi_expct},
\be
\lp^\alpha\,
\sqrt{\gn}
\bra{g}\hat{\Phi}\ket{g}
&\!\!=\!\!&
-\sqrt{\frac{\lp}{\mpl}}\,
\int_0^\infty
\frac{k^{2}\,\d k}{2\,\pi^{2}}\,\sqrt{\frac{2\,\lp\,\mpl}{k}}\, 
\frac{4\,\pi^2\,f(t)}{\sqrt{2\,k}\,\lp\,i^\alpha}\,\delta^{(1+\alpha)}(k)\,
j_0(k\,r)
\nonumber
\\
&\!\!=\!\!&
-2\,f(t)
\int_0^\infty
\frac{\d k}{i^\alpha\,r}\,\delta^{(1+\alpha)}(k)\,\sin(k\,r)
\nonumber
\\
&\!\!=\!\!&
f(t)\,r^\alpha
\int_{-\infty}^{+\infty}
{\d k}\,\delta(k)\,\cos(k\,r)
\nonumber
\\
&\!\!=\!\!&
f(t)\,r^\alpha
\ ,
\ee
for $\alpha=0$ or $\alpha=2$.
\par
From the Gaussian regularisation~\eqref{deltaReg}, we find,
for $\alpha=0$,
\par
\be
\sqrt{\gn}
\bra{g}\hat{\Phi}\ket{g}
&\!\!=\!\!&
\sqrt{\frac{\lp}{\mpl}}\,
\int_0^\infty
\frac{k^{2}\,\d k}{2\,\pi^{2}}\,
\sqrt{\frac{2\,\lp\,\mpl}{k}}
\,\frac{4\,\pi^2\,f(t)}{\sqrt{2\,k}\,\lp}\,
\frac{2\,k\,e^{-\frac{k^2}{\sigma^2}}}{\sqrt{\pi}\,\sigma^3}
\,j_0(k\,r)
\nonumber
\\
&\!\!=\!\!&
\frac{4\,f(t)}{r}
\int_0^\infty
\frac{k\,\d k}{\sqrt{\pi}\,\sigma^3}\,
e^{-\frac{k^2}{\sigma^2}}\,
\sin(k\,r)
\nonumber
\\
&\!\!=\!\!&
e^{-\sigma^2\,r^2/4}\,f(t)
\ ,
\label{eq:regQC1}
\ee
and, for $\alpha=2$,
\be
\lp^2\,
\sqrt{\gn}
\bra{g}\hat{\Phi}\ket{g}
&\!\!=\!\!&
\int_0^\infty
\frac{\d k \, k^2}{2\,\pi^{2}}\,
\frac{4\,\pi^2}{i^2\,k}\,
\frac{4\,k\,(2\,k^2-3\,\sigma^2)}{\sqrt{\pi}\,\sigma^7}\,
e^{-\frac{k^2}{\sigma^2}}\,
f(t)\,
j_{0}(kr)
\nonumber
\\
&\!\!=\!\!&
-\frac{8\,f(t)}{\sqrt{\pi}\, r \,\sigma^7}
\int_{0}^{\infty}
\d k \, k \,
(2\,k^2-3\, \sigma^2)\,
e^{-\frac{k^2}{\sigma^2}}\,
\sin(kr)
\nonumber
\\
&\!\!=\!\!&
e^{-\frac{\sigma^2\,r^2}{4}}\,
f(t)\,
r^2
\ .
\label{eq:regQC2}
\ee
\begin{figure}[t]
  \centering
   \includegraphics[width=0.5\linewidth]{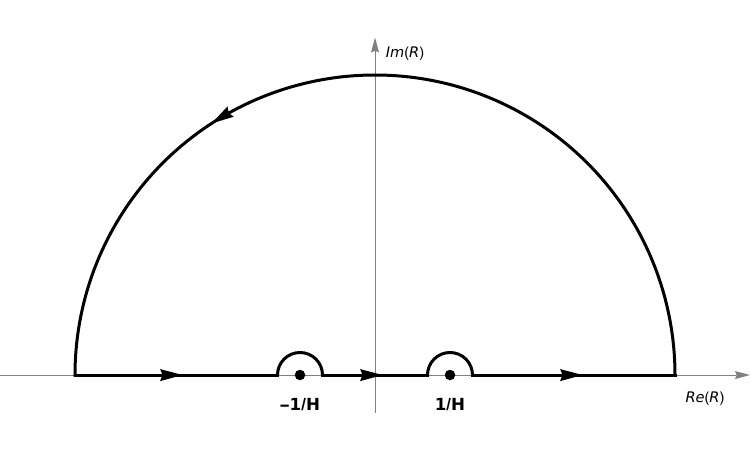}
  \caption{Contour used in the integration of the principal value.}
  \label{fig:cont}
\end{figure}
\section{Static patch in spherical coordinates}
\setcounter{equation}{0}
\label{A:useful2}
The Fourier transform for the $g_{RR}$ component is
\be  
\label{eq:fkRR}
\tilde f_k(T)
&\!\!=\!\!&
4\,\pi
\int_0^\infty
R^2\,\d R\,j_0(k\,R)\,
\frac{e^{i\,k\,T}}{(1-H^2\,R^2)}
\nonumber
\\
&\!\!=\!\!&
\frac{4\,\pi}{k}\,
e^{i\,k\,T}\,
\int_{0}^{\infty}
\frac{R\,\d R}{1-H^2\,R^2}\,
\sin(k\,R)
\nonumber
\\
&\!\!=\!\!&
\frac{2\,\pi}{i\,k}\,
e^{i\,k\,T}\,
\int_{-\infty}^{+\infty}
\frac{R\,\d R}{1-H^2\,R^2} \,e^{i\,k\,R}
\ .
\ee
The integrand has two simple poles at $R=\pm H^{-1}$ and a non-vanishing
Cauchy principal value ($\rm p.v.$) which can be computed from the contour
integral on the complex plane displayed in Fig.~\ref{fig:cont}, 
\be
\label{eq:pv}
\mathrm{p.v.}
\int_{-\infty}^{+\infty}
\frac{R\,\d R }{1-H^2\,R^2} \,e^{i\,k\,R}
=
-\frac{i\,\pi}{H^2}\, \cos(k/H)
\ ,
\ee
from which we obtain
\be
\tilde f_k(T)
=
-
\frac{2\,\pi^2}{k\,H^2}\,
\cos(k/H)\,
e^{i\,k\,T}
\ ,
\label{fkT}
\ee
or $\gamma_k=k\,T$ and 
\be
g_k
=
-
\frac{2\,\pi^2}{\sqrt{2\,k}\,H^2\,\lp}
\,\cos(k/H)
\ .
\ee
\par
The quantum-corrected metric function $\bra{g}\hat g_{TT}\ket{g}$ is now given by
\be
\sqrt{\gn}
\bra{g}\hat{\Phi}_{TT}\ket{g}
&\!\!=\!\!&
\int_0^\infty
\frac{k^{2}\,\d k}{2\,\pi^{2}}\,
\frac{4\,\pi^2}{k}
\left[
  \frac{2\,k}{\sqrt{\pi}\,\Sigma }
  +
  H^2\,
  \frac{4\,k\,(2\,k^2-3\,\Sigma^2)}{\sqrt{\pi}\,\Sigma^7}
\right]
e^{-\frac{k^2}{\Sigma^2}}\,
j_0(k\,R)
\nonumber
\\
&\!\!=\!\!&
\frac{4}{\sqrt{\pi}\, R \,\Sigma^7}\,
\int_{0}^{\infty}
k\,\d k
\left(
  4\,H^2\,k^2+\Sigma^4-6\,H^2\,\Sigma^2
\right)
e^{-\frac{k^2}{\Sigma^2}}\,
\sin(kR)
\nonumber
\\
&\!\!=\!\!&
e^{-\frac{R^2\,\Sigma^2}{4}}\,(1-H^2\,R^2)
\ .
\label{eq:gttQCstatic}
\ee
Likewise, for the metric component $\bra{g}\hat g_{RR}\ket{g}$ we obtain
\be
\sqrt{\gn}
\bra{g}\hat{\Phi}_{RR}\ket{g}
&\!\!=\!\!&
-
\frac{1}{H^2\,R}
\int_0^\infty
\d k\,
\sin(k\,R)\,
\cos(k/H)\,
e^{-\frac{\ell^2\,k^2}{4}}
\nonumber
\\
&\!\!=\!\!&
-\frac{1}{H^2\,\ell\,R}
\left[
  D_+\!\left(\frac{H\,R-1}{H\,\ell}\right)
  +
  D_+\!\left(\frac{H\,R+1}{H\,\ell}\right)
\right]
\ ,
\label{eq:grrQCgauss}
\ee
where 
\be
D_+(z)
=
e^{-z^2}
\int_{0}^{z}
\d t \,e^{t^2}
\label{dawson}
\ee
is the Dawson function.

In the coherent state $|g\rangle$ which gives $\bra{g}\hat \Phi_{RR}\ket{g}$, the normalisation
reads
\be
\mathcal{N}
&\!\!=\!\!&
\int_{0}^{\infty}
\frac{k^3\,\d k}{4\,\pi^2\,\lp^2}\,
\frac{(2\,\pi^2)^2}{k^2\,H^4}\,
\cos^2\left(\frac{k}{H}\right)
\nonumber
\\
&\!\!=\!\!&
\frac{\pi^2}{2\,H^4\,\ell_p^2}
\int_{0}^{\infty}
k\,\d k
\left[
  1+\cos\left(\frac{2\,k}{H}\right)
\right]
\ ,
\label{NgRR} 
\ee
which diverges in the UV, for $k \to +\infty$.
A coherent state with finite occupation number can be obtained by introducing a UV length scale
$\ell$ which regularises the coefficients
\be
g_k
\to  
g_k\,e^{-\frac{\ell^2\,k^2}{4}}
\ .
\ee
The normalisation factor in Eq.~\eqref{NgRR} now becomes
\be
\mathcal{N}
&\!\!=\!\!&
\frac{\pi^4}{2\,H^4\,\ell_p^2}
\int_{0}^{\infty}
k\,\d k
\left[
  1+\cos\left(\frac{2\,k}{H}\right)
\right]
e^{-\frac{\ell^2\,k^2}{2}}
\nonumber
\\
&\!\!=\!\!&
\frac{\pi^4}{H^4\,\ell_p^2\,\ell^2}
\left[
  1
  -
  \frac{\sqrt{2}}{H\,\ell}\,\, D_+\!\left(\frac{\sqrt{2}}{H\,\ell}\right)
\right]
\ ,
\ee
where $D_+$ is the Dawson function~\eqref{dawson}.
For $H\,\ell\ll 1$, we then find
\be
\mathcal{N}
\simeq
\frac{\pi^4}{H^4\,\ell_p^2\,\ell^2}
\ .
\ee
\end{document}